# Rapidly reconfigurable radio-frequency arbitrary waveforms synthesized on a CMOS photonic chip

Jian Wang[1,†], Hao Shen[1,†], Li Fan[1], Rui Wu[1], Ben Niu[1], Leo T. Varghese[1], Yi Xuan[1], Daniel E. Leaird[1], Xi Wang[2], Fuwan Gan[2]\*, Andrew M. Weiner[1]\*, and Minghao Qi[1, 2]\*

[1]School of Electrical and Computer Engineering and Birck Nanotechnology Centre, Purdue University, 1205 W. State Street, West Lafayette, IN 47907, USA

[2]State Key Laboratory of Functional Materials for Informatics, Shanghai Institute of Microsystem and Information Technology, Chinese Academy of Sciences, 865 Changning Road, Shanghai 200050, China

[†]These authors contributed equally to this work.

[*]e-mail: mqi@purdue.edu, amw@purdue.edu, fuwan@mail.sim.ac.cn

**Photonic methods of radio-frequency waveform generation and processing provide performance and flexibility over electronic methods due to the ultrawide bandwidth offered by the optical carriers. However, they suffer from lack of integration and slow reconfiguration speed. Here we propose an architecture of integrated photonic RF waveform generation and processing, and implement it on a silicon chip fabricated in a semiconductor manufacturing foundry. Our device can generate programmable RF bursts or continuous waveforms with only the light source, electrical drives/controls and detectors being off chip. It turns on and off an individual pulse in the RF burst within 4 nanoseconds, achieving a reconfiguration speed three orders of magnitude faster than thermal tuning. The on-chip optical**

**delay elements offers an integrated approach to accurately manipulate individual RF waveform features without constrains set by the speed and timing jitter of electronics, and should find broad applications ranging from high-speed wireless to defense electronics.**

Arbitrary waveform generation (AWG) is critical to many radio-frequency (RF) and time-domain tests, measurements and applications, including ultrawide-bandwidth and multiple-access wireless communications, electronic counter-measures and radar systems. Commercial electronic methods of AWG have high output precision as well as versatility, but limited sampling rates and analog bandwidths, primarily due to timing jitter and the speed and linearity of digital-to-analog converters (DACs). Photonic methods of generating and processing radio-frequency signals have offered bandwidths and intra-pulse modulation that are unattainable by electronic means[1-4]. However, with commercial electronic AWG improving in speed and timing, photonic AWG must increase the reconfiguration speed, duty cycle and system stability, as well as reduce size, weight, power consumption and cost.

Photonics assisted RF generation typically involves the generation of delayed and scaled replicas of RF signals or impulses in the optical domain with subsequent recombination to achieve the desired RF output (Figure 1**a**). Conventionally, multiple optical carriers at different wavelengths go through a pulse shaper[5] to acquire desired amplitude and/or phase changes. We have first demonstrated an on-chip silicon multi-channel reconfigurable microring filter bank based pulse shaper that generated 8 replicas of RF impulses at 8 tunable optical frequencies from a mode-locked ultra-fast optical pulse, scaling them to specified amplitudes before recombination[6]. However, to achieve appropriate delays between the 8 replicas, the recombined signals needed to go through a dispersive device, such as a long spool of optical fiber (5.5 km long[6]) or linearly chirped fiber Bragg grating (FBG)[7]. Such a process is called frequency-to-time mapping[1,2], which provides accurate and continuous adjustment of delays, a feature difficult to achieve by electronic means if delay differences between adjacent carrier wavelengths are as short as a few tens of picoseconds. Unfortunately, dispersive devices are not amenable to miniaturization and integration. For example, to achieve the same amount of dispersion as the 5.5km-long fiber, a 21 meter long silicon waveguide with a dispersion coefficient of 4400 ps/(km nm) is required[8]. The resulting high propagation loss and large footprint (cm$^2$) make this option impractical. Chirped Bragg gratings are capable of providing larger dispersion coefficients but are further constrained by the limited delay-bandwidth products for on-chip implementations[9,10]. Photonic crystals show good performance in delaying optical signals but also impose significant challenges in fabrication[11-13].

## Results

**Time-domain waveform synthesis using silicon optical delay lines.** Here we introduce an architecture of RF AWG based on a direct time-domain synthesis method which no longer requires a highly dispersive and physically long medium. Figure 1**a** depicts the schematic, where a mode-locked femtosecond pulse is tailored by a multi-channel (eight in Figure 1**a**) filter bank and a delay element array in both frequency and time domains. First, a ultra-fast pulse (~130 fs, ~80 nm bandwidth) is sampled by 8 microring filters that download and generate 8 replica pulses at different frequencies (denoted by their resonance wavelengths $\lambda_i$). The wide bandwidth of the input pulse covers 6 free spectral ranges (FSRs, ~16 nm) of the 5μm-radius silicon microrings, and the frequency components throughout all the 6 FSRs coherently add up to form the channel replica pulses. Since each replica generation microring can only download part of the input pulse's spectrum to one channel, the temporal width of each replica experience pulse broadening, which is determined by both the optical round-trip time in the ring (group index $n_g \approx 4.25$ for the TM mode) and the ring's quality factor ($Q$). To avoid excessive broadening of pulses as well as to reduce the insertion loss of the pulse shaper, the microrings are designed to over-couple to bus waveguides (with a power coupling coefficient of $\kappa^2 \approx 0.1$). The channel frequency spacings in the filter bank can be arbitrary and non-uniform as long as they are not too small to cause interference between channels. In our case an approximately uniform channel spacing of 1 nm is chosen (with a designed radius step increase of 8 nm) but no effort is put in to accurately achieve it. This reduces the challenge in fabrication.

The key component of the photonic AWG is the on-chip tunable delay element. Each optical delay line consists of a fixed part and a tunable part (Figure 1**b**). The fixed part is a long waveguide with a delay value at multiples of 25 ps, and each channel has 25 ps more delay than the previous one. The tunable part consists of 41 cascaded all-pass ring resonators whose radii vary with a 2-nm interval and with the radius of the central one same as that of the replica generation ring in the same channel. All the microrings in the delay line are designed in the over-coupling regime (with a power coupling coefficient of $\kappa^2 \approx 0.6$) to migitate dispersion induced pulse distortion as well as to reduce the insertion loss[14,15]. According to our simulation the delay spectrum of the microring array is not significantly affected by random errors in microring's radii introduced by the electron-beam lithography or optical lithoghraphy, which offers considerable fabrication tolerance. Serpentine micro-heaters cover the entire tunable delay line area, tuning the resonances of all 41 rings and shifting the delay spectrum through the thermo-optic effect (see the inset of Figure 1a). In the experiment, a tunable delay range up to 27 ps, greater than the fixed delay step of 25 ps across adjacent channels, was observed (Figure 1c). The slight change in amplitude is a consequence of delay, caused by additional propagation loss when light propagates longer within the ring array.

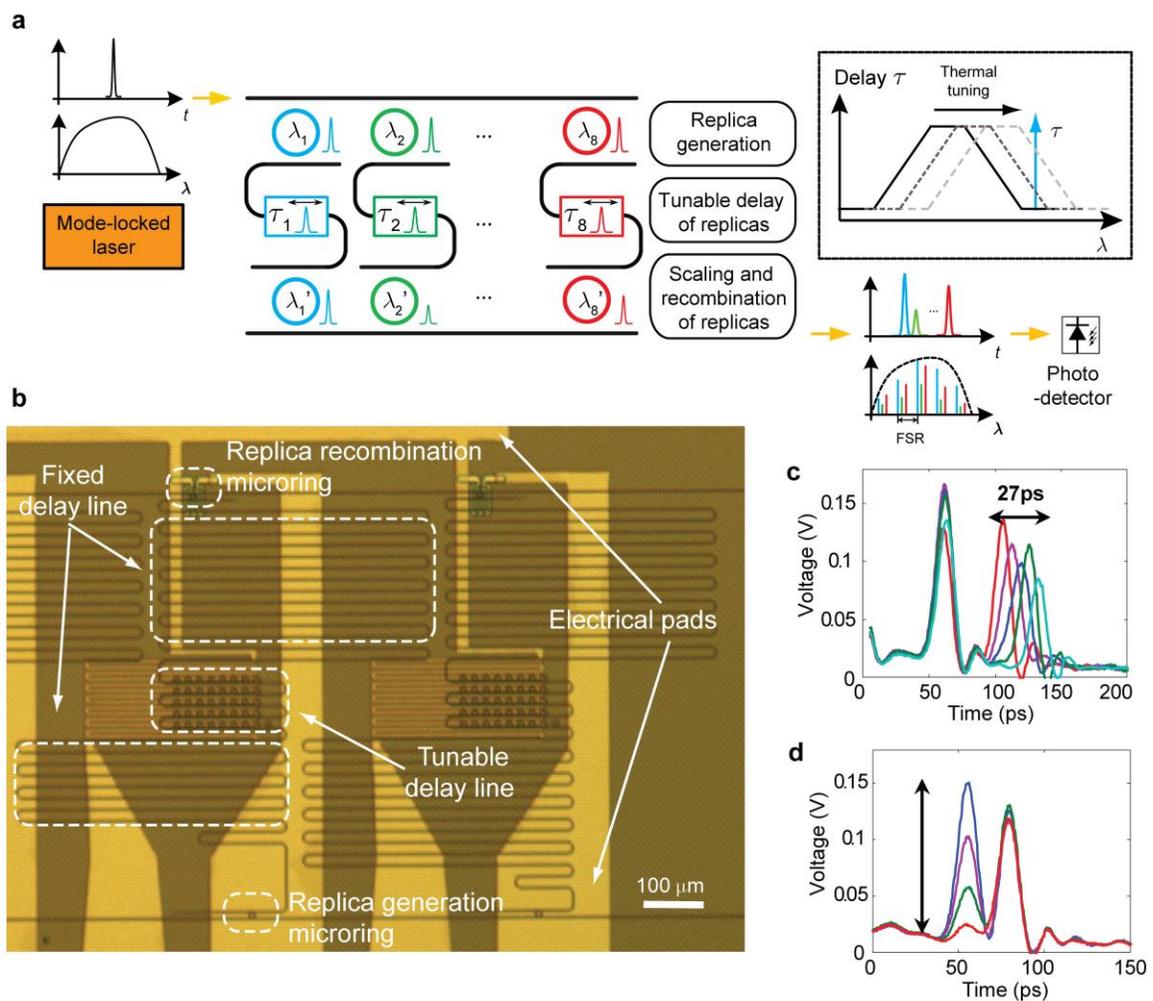

**Figure 1 | Mechanism and tunability of the time-domain synthesis method of arbitrary RF waveform.** (a) A schematic of the time-domain synthesis method of RF AWG using a silicon pulse shaper. A mode-locked femtosecond laser is

used as the light source, and the RF waveforms are obtained through off-chip optical/electrical conversion of the shaped optical waveforms using a photo-detector. FSR: free spectral range of a microring. The inset shows a schematic of the thermally tuned optical delay spectrum of cascaded all-pass filters in the tunable delay line. (b) An optical image of optical delay-lines and add-drop filters in a fabricated device. (c) and (d) The delay and amplitude of an individual pulse is thermally tuned without appreciably affecting the pulse from an adjacent channel.

The amplitude of the pulse in each channel can be controlled by mismatching the resonance wavelength $\lambda_i$' of the replica recombination ring from that of the replica generation ring in the same channel through the thermo-optic effect (TOE). The perimeter of the replica recombination ring is designed to be slightly smaller than that of the replica generation ring, and micro-heaters are placed on top of each recombination ring only. Figure 1**d** shows almost 100% amplitude control without significantly affecting the reference pulse in the adjacent channel.

**Synthesis of RF bursts.** With the capabilities of independent amplitude and delay control, RF waveforms with sophisticated amplitude, frequency and phase modulations were created (Figure 2). The first example of a 40 GHz RF burst (Figure 2**a**) was created with a preset channel delay spacing of 25 ps by the fixed delay lines and with slight tunable delay applied for minor adjustment. By apodizing the envelop of the 40 GHz burst to a smooth shape (Figure 1**c**), the mainlobe-sidelobe suppression ratio in the corresponding RF spectrum was improved from 8 dB (Figure 2**b**) to 14 dB (Figure 2**d**). The next three examples are a single-tone 30 GHz RF burst, a 40 GHz RF burst with an abrupt π phase shift, and a frequency-chirped signal (Figures 2**e**, 2**g**, and 2**h**, respectively). The chirped signal shows a frequency modulation from 30 GHz to 50 GHz, containing very broadband frequency components, which can be useful in chirped radar systems. The time-domain approach produces ultrabroadband RF arbitrary waveforms with performance comparable to that which we previously demonstrated using a Si-photonic pulse shaper chip based on the frequency domain approach based on frequency-to-time mapping[6], but without the need for an external multi-kilometer length of optical fiber.

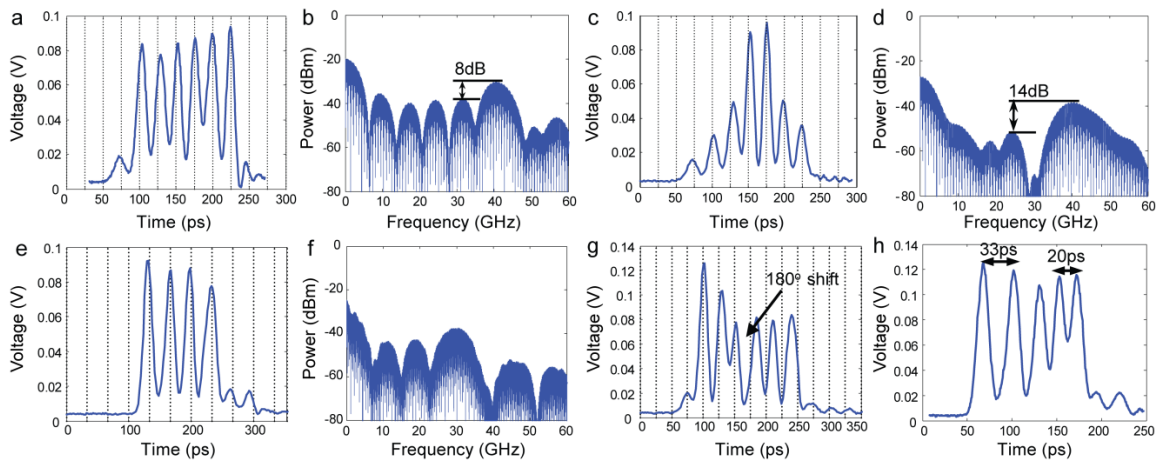

**Figure 2 | Various RF bursts and their spectra calculated via Fourier transform.** (a) and (b) A 40 GHz RF burst and the calculated spectrum, where the mainlobe-sidelobe suppression ratio is ~8 dB. (c) and (d) An apodized 40 GHz RF burst and the calculated spectrum, where the pulse apodization increases the mainlobe-sidelobe suppression ratio to ~14 dB. (e) and (f) A 30 GHz RF burst and its calculated spectrum. (g) π phase shift in a 40 GHz RF burst. (h) An up-chirping signal with frequencies ranging from 30 GHz to 50 GHz.

**Synthesis of continuous RF waveforms.** The generated RF waveforms so far are in the form of repetitive RF bursts, each synchronized with a particular pulse of the low-repetition-rate input femtosecond pulse laser. The duty cycle is also low. These RF waveforms are inefficient in communications and information processing where high data throughput is paramount. To obtain RF waveforms with high duty cycles, one can either use a laser source with a higher repetition rate that is close to the inverse of the temporal width of the bursts or introduce more features in one RF burst. Here we adopted the first scheme and chose a 10 GHz optical frequency comb as a new light source to generate continuous RF waveforms[16]. The optical frequency comb is generated from a CW laser source based on periodic electro-optic modulation[17]. The carrier wavelength of 1542 nm of the comb coincides with the wavelength range of the pulse shaper, and the comb spans over a 6 nm wavelength range covering more than four channels of the shaper. Figure 3**a** shows an example of a 40 GHz continuous RF waveform where four channels were selected with the channel delay spacing preset at 25 ps. A 30 GHz continuous waveform was created when three channels were selected with the channel delay spacing adjusted to 33 ps through tunable delay (Figure 3**b**). Therefore, single-tone continuous RF waveforms can be generated using a high-repetition-rate light source and the photonic chip with properly preset channel delay.

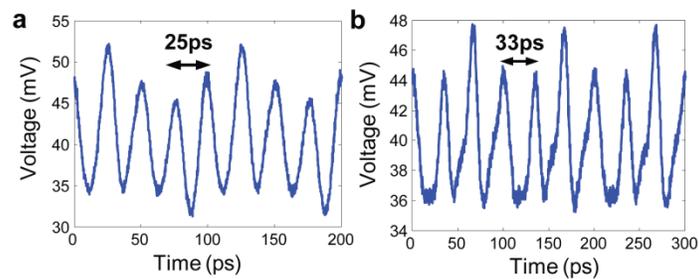

**Figure 3 | Single-tone continuous RF waveform generation using a 10 GHz optical frequency comb source.** (a) A 40 GHz continuous RF waveform. (b) A 30 GHz continuous RF waveform.

**Rapidly reconfigurable waveform enabled by a silicon modulator.** Truly arbitrary RF waveforms allow modulation of features in a pulse by pulse fashion, which can take the full advantage of the wide bandwidth of RF waveforms. All the above-mentioned reconfiguration of the waveform generator is achieved through the thermo-optic effect in silicon, which has a response time around tens of micro-seconds[18]. While this is already faster than bulk pulse shapers employing liquid crystal modulators, in practice fast reconfiguration time close to the inverse of the waveform repetition rate is desirable for many applications such as wireless data communications. An advantage of silicon over liquid crystals or other materials in photonic integrated circuits, such as $SiO_2$, is that it offers the strong free-carrier effect with a sub-nanosecond response time[19]. Here we employ a free-carrier depletion based Mach-Zehnder intensity modulator to demonstrate rapid control of an individual pulse amplitude [20-23].

Figure 4**a** is an optical image of the silicon chip for rapidly reconfigurable RF AWG, which is fabricated in a CMOS foundry using 0.13 μm optical lithography. The silicon chip consists of an 8-channel pulse shaper and an intensity modulator embedded in the 8$^{th}$ channel of the shaper. For proof of concept, only one modulator is incorporated. Since the modulator is designed for TE-polarized light, all the microrings in the shaper need to work in the over-coupling regime for TE light, which requires a narrow coupling gap of less than 100 nm in a 5μm-radius microring. For optical lithography capable of defining 200nm-wide gaps, a racetrack microring with a longer coupling distance with bus waveguide can achieve effective over-coupling with a 200nm-wide gap (inset of Figure 4**a**). We first generated 16 GHz RF bursts by selecting 4 channels in the shaper without fast modulation. With a 250 MHz optical frequency comb source, the amplitude of each pulse of the created waveform can be controlled completely and individually without appreciably affecting other pulses (Figure 4**c**). Next, we modulated the intensity of the pulse from the last channel. The experimental setup is shown in Figure 4**b**. Two consecutive RF waveforms separated by ~4 ns clearly show the fast reconfiguration of the last channel pulse (Figure 4**d**). The modulation depth of ~3:1 is consistent with that measured with CW light.

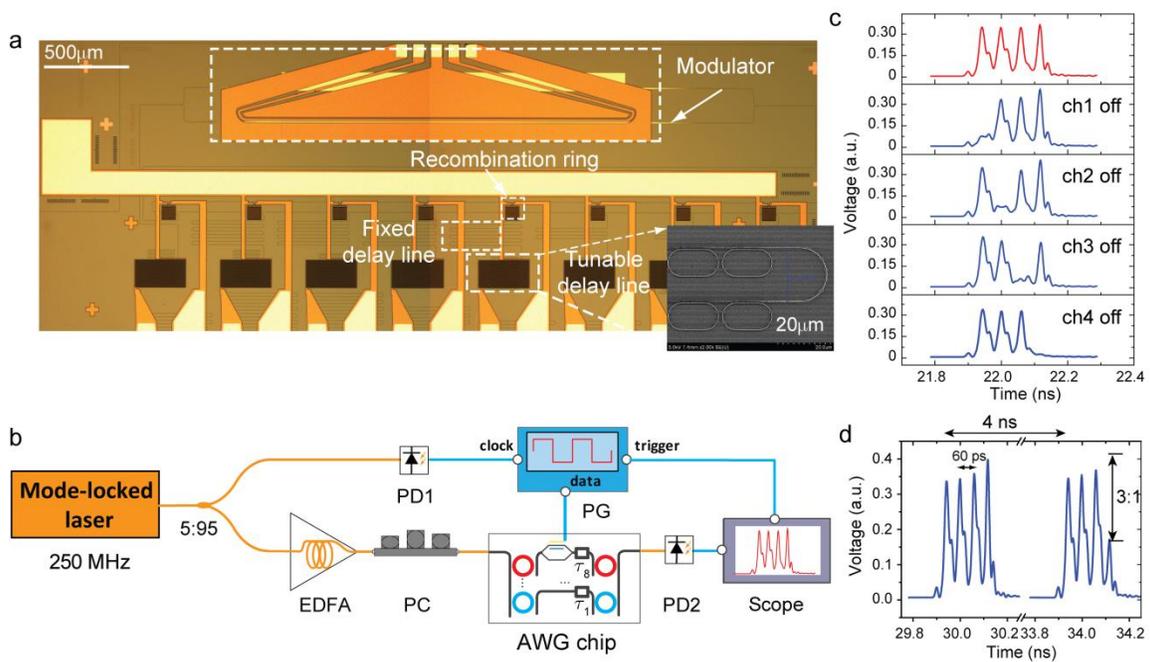

**Figure** 4 **| Rapidly reconfigurable AWG with synchronized amplitude modulation through a silicon electro-optic modulator embedded in the shaper.** (a) An optical image of a silicon chip for the nanosecond-reconfigurable waveform generation, which consists of an 8-channel silicon pulse shaper and a 2mm-long electro-optic modulator embedded in the 8$^{th}$ channel of the shaper. The inset is an SEM picture of a tunable delay-line that consists of cascaded racetrack microrings. (b) Experimental setup. PD: photo-detector; PG: pattern

generator; EDFA: erbium-doped fiber amplifier; PC: polarization controller. (c) Amplitude control of individual pulses through resonance mismatching enabled by the thermo-optic effect. (d) Two consecutive RF bursts show the effect of amplitude modulation using the silicon intensity modulator.

The updating speed is currently limited by the repetition rate of the comb laser source as well as the design of the modulator embedded in the shaper. 44 Gbps modulation has been demonstrated using the same fabrication process[24], so in principle the reconfiguration time of RF waveforms can reach the deep sub-nanosecond range and continuous RF waveforms with intensity modulation at any pulse can be achieved. For a single-tone RF burst, our scheme allows the modulators to be run at a speed reduced by a factor equal to the number of peaks per burst, with significantly relaxed timing requirements. Rapid tuning of delay is also necessary. This could be achieved by directly modulating all all-pass filters in a tunable delay line that share a common RF driving signal. Recently a cascaded-ring modulator has been demonstrated with a 0.2 nm bandwidth, allowing small distortion to the RF signals[25].

**Discussion**

A dispersive device, such as a spool of optical fiber or a fiber Bragg grating, has been one of the central elements that allow photonics to process microwave signals. Therefore it is almost ubiquitous in microwave photonic processing systems. However, direct integration of this element onto a chip will lead to prohibitively long waveguides and/or intolerably high insertion loss. Our approach of controlling the optical delay of each individual frequency component bypasses the long dispersive element. Functionally, our approach offers discrete frequency-to-time mapping, with the channel wavelength no longer tied to a specific delay, an advantage that offers additional flexibility. It does lose the capability of continuous tuning of waveform timing features, but that can be mitigated by reducing the digitization error, *i.e.* increasing the number of channels which is much easier to achieve in the integrated form than in bulk optics. Meanwhile, the demonstration of integrating modulators with pulse shapers on the same chip may also have impact for on-chip all-optical RF processing. Our approach should be applicable to a variety of microwave photonic applications beyond the arbitrary waveform generation application considered here, including RF photonic filters[26-28], photonic beamformers for phased array receive antennas[29,30], and precompensation of dispersive microwave antenna links[31]. The small footprint (4 mm by 2.7 mm for an 8-channel shaper) and integrated nature of our device allow multiple copies of devices to be fabricated on the same chip, potentially enabling parallel processing of the same input RF signal at different bands of interest, *i.e.* functioning as an RF channeller.

   The time-domain synthesis scheme we demonstrate here allows RF arbitray waveform generation with amplitude, frequency, and phase control with performance comparable to that achieved in our previous Si photonic shaper chip designed according to a frequency-to-time mapping approach[6]. Moreover, the integration of fast silicon modulators allows agility, or speed of reconfiguration, unachievable in previous demonstrations using either bulk optical pulse shapers or photonic integrated circuits based on silica or silicon nitride. Such rapidly reconfigurable RF waveforms generated through integrated modulators enable pulse-by-pulse variable arbitrary RF signal generation that is highly desirable for radar applications. Continuous waveforms we demonstrated extend the AWG applications to high data throughput applications, and together with high updating speeds could potentially match the flexibility of electronic AWG while maintaining the bandwidth and speed advantages of photonics. For the remaining RF-AWG elements, the photo-detector can be readily integrated[32] and with recent coherent comb generation in $Si_3N_4$ microresonators[33,34], the light source might also be integrated to achieve a fully monolithic solution of AWG, with potential to better electronic AWG in bandwidth, reconfiguration speed, timing jitter, size, and power consumption.

**Methods**

**Design and fabrication of the preliminary chip (Figure 1).** The waveguides in the preliminary pulse shaper fabricated at Purdue have a cross-section of 500 nm × 250 nm (width × height). The radius of each replica generation or recombination microring is around 5 μm, with a step increase of 8 nm to separate its resonance from that in the previous channel by ~1 nm. In each frequency channel, the tunable delay line contains 41 all-pass microrings in strong coupling regime (with a power coupling coefficient of $\kappa^2 = 0.6$). The preliminary chip was fabricated on silicon-on-insulator (SOI) wafer with high resolution electron-beam lithography (EBL). After the silicon waveguide and resonators were defined, a 2μm-thick silicon dioxide layer was deposited over the device as a buffered layer, using plasma enhanced chemical vapor deposition (PECVD). On the oxide layer, electrical heaters were defined by optical lithography. The compact heaters on top of the microrings were made of titanium and the serpentine long heaters on top of the tunable delay lines were made of aluminum together with the contact leads.

**Measurement of the preliminary chip.** In the continuous RF waveform measurement, the original comb generated based on electro-optic modulation spans over a wavelength range of 2.4 nm, which is not sufficient for the AWG application. We broadened the comb bandwidth to 6 nm via nonlinear propagation in a length of dispersion decreasing fiber (DDF)[35], which can cover more than 4 channels of the shaper. Before the DDF, a low noise EDFA and a high gain EDFA were used sequentially to boost the comb signal to a power level sufficient for the adiabatic pulse compression.

**Fabrication and measurement of the foundry chip (Figure 4).** The device was fabricated in a commercial CMOS foundry using 0.13 μm technology on 8'' SOI wafers with a 220nm-thick top silicon layer and a layer of 2μm-thick buried dioxide. We first characterized the foundry chip with a tunable CW light source and found that the drop-port transmission in the $8^{th}$ channel of the shaper was 10 dB lower than all other channels due to the insertion loss of the modulator. To show the modulation of AWG more clearly, we compensated this loss by pre-attenuating all frequency components outside the $8^{th}$ channel in the input pulses by 10 dB using a commercial pulse shaper throughout a wavelength range from 1528 nm to 1568 nm. The experimental setup of the rapidly reconfigurable AWG is shown in Figure 4**b**, where 5% of the input pulse power was converted to an electrical signal serving as a clock signal to the pattern generator (PG), which drove the intensity modulator working synchronously with the RF waveform. An RF amplifier (10-1500 MHz) was used to boost the clock signal, which was followed by an RF low-pass filter (DC-400 MHz) to filter out the high order harmonics. In the experiment, a 250 MHz frequency comb (Menlo Systems) was chosen as the light source with a repetition rate high enough to drive the pattern generator. The pulse width was compressed to ~400 fs using a length of ~12.5m-long dispersion compensation fiber (DCF). A post-device EDFA was used to boost the optical waveform output from chip and a 12nm 3dB-bandwidth optical band-pass filter was used afterwards to remove the strong atomic spontaneous emission (ASE) around the 1530 nm wavelength range from the EDFA.

**Acknowledgements**

Jian Wang acknowledges helpful discussion with Jing Wang. This work was supported by Air Force Office of Scientific Research grant FA9550-08-1-0379, National Science Foundation grants ECCS-0925759, ECCS-0901383 and CNS-1126688, Defense Threat Reduction Agency grant HDTRA1-10-1-0106, and National Institute of Health grant 1R01RR026273-01. This work was also supported in part by the Office of the Assistant Secretary of Defense for Research and Engineering under the National Security Science and Engineering Faculty Fellowship program through grant N00244-09-1-0068 from the Naval Postgraduate School.

## Author contributions

H. S., J. W. and M. Q. designed the devices. L.F. and L.T.V. fabricated the preliminary devices. H. S., J. W., and R. W. measured the devices. F.G. designed the on-chip modulator. Y. X. and B. N. helped in the fabrication of preliminary devices. D. E. L helped in the measurement. M. Q. and A. M. W. conceived the idea and supervised the investigation. J. W., M. Q., A. M. W., and H. S. wrote the manuscript. All discussed the results and commented on the manuscript.

## Additional Information

**Competing financial interests**: The authors declare no competing financial interests.

**Reprints and permission information** is available online at http://www.nature.com/reprints.

Correspondence and requests for materials should be addressed to M. Q., A. M. W. or F. G..